\documentstyle[preprint,aps]{revtex}

\begin{document}
\draft
\preprint{KANAZAWA 93-09}
\title{
Monopole Condensation and Confinement in SU(2) QCD (1)
}
\author{
Hiroshi Shiba\cite{sh} and Tsuneo Suzuki\cite{su}}
\address{
Department of Physics, Kanazawa University, Kanazawa 920-11, Japan
}
\date{\today}
\maketitle
\begin{abstract}
An effective monopole action is derived from vacuum configurations 
after abelian projection in the maximally abelian gauge in $SU(2)$ QCD. 
Entropy dominance over energy of monopole loops is seen 
on the renormalized lattice with the spacing 
$b>b_c\simeq 5.2\times10^{-3}
\Lambda_L^{-1}$ when the physical volume of the system 
is large enough.
QCD confinement may be 
interpreted as the (dual) Meissner effect 
due to the monopole condensation.
\end{abstract}


\narrowtext

To understand confinement mechanism is very important 
but still unresolved 
problem in particle physics. 
A promising idea is that the (dual) 
Meissner effect due to condensation 
of some magnetic
quantity is the color confinement mechanism 
in QCD\cite{thooft1,mandel}.
This picture is realized in the confinement phase of lattice compact
QED\cite{poly,bank,degrand}. Especially interesting 
are the following facts.

1) A dual transformation can be done, leading us to 
an action describing a monopole Coulomb gas.
\cite{bank,peskin,frolich,smit}. Monopole condensation is shown to 
occur in the confinement phase from energy-entropy balance. 

2) The monopole contribution alone can reproduce the full value
of the string tension\cite{stack}.

In the case of QCD, there is a difficult problem. We have to find a 
color magnetic quantity in QCD. In this respect, 
the 'tHooft idea of abelian
projection of QCD\cite{thooft2} is very interesting.  
The abelian projection of QCD is to perform a partial gauge-fixing  
such that the maximal abelian torus group remains unbroken. Then 
QCD can be regarded as a $U(1)\times U(1)$ abelian gauge theory
with magnetic monopoles and electric charges. 't Hooft 
conjectured that the condensation 
of the abelian monopoles is the confinement mechanism in 
QCD\cite{thooft2}.
There are, however, infinite ways of extracting such an 
abelian theory out of $SU(3)$ QCD. It seems important to find a 
good gauge in which the conjecture is seen to be realized.

A gauge called maximally abelian (MA) gauge has been shown to be very 
interesting \cite{kron,yotsu,suzu92,suzu93}. In the MA gauge, 
there are phenomena which may be called abelian dominance
\cite{yotsu,hio}. 
Moreover the monopole current 
$k_{\mu}(s)$ can be 
defined similarly as in compact QED\cite{degrand}.
It is shown in the MA gauge that the abelian monopoles
are dense and dynamical in the confinement phase, whereas they
are dilute and static in the deconfinement phase\cite{suzu93}.

The abelian dominance suggests that a set 
of $U(1)$ invariant operators 
are enough to describe confinement in the MA gauge. 
Then there must exist an effective $U(1)$ action 
describing confinement.
Let us try  to derive an effective $U(1)$ action on the dual 
lattice as done in compact QED from monopole distributions given by 
Monte-Carlo simulations.

Swendsen\cite{swendsen} developed a method of 
determining an action from 
a given ensemble of configurations. The method 
can be applied to our case.
A theory of monopole loops is given in general by 
the following partition 
function  
$$ Z= (\prod_{s,\mu}\sum_{k_{\mu}(s)=-\infty}^{\infty})
(\prod_s \delta_{\partial'_{\mu}k_{\mu}(s),0})
\exp(-\sum_i f_i S_i [k]), $$
where $\partial'$ is a backward derivative on a dual lattice and $k_{\mu}(s)$ 
is the conserved integer-valued monopole current. $f_i$ is a coupling constant 
of an interaction $S_i [k]$. 
Since the dynamical variables here are $k_{\mu}(s)$ satisfying the 
conservation rule, it is necessary to extend the 
original Swendsen method by 
considering a plaquette $(s',\mu',\nu')$ instead of a link.
Introducing a new set of coupling constants 
$ \{\tilde{f}_i\} $, define
\begin{equation}
\bar{S}_i [k]= \frac{\sum_{M=-\infty}^{\infty}S_i[k']
\exp(-\sum_i \tilde{f}_i S_i [k'])}
{\exp(-\sum_i \tilde{f}_i S_i [k'])},
\label{sbar}
\end{equation}
where 
$k'_{\mu}(s)  =  k_{\mu}(s) + M(\delta_{s,s'}\delta_{\mu\mu'}
+ \delta_{s,s'+\hat{\mu}'}\delta_{\mu\nu'}  
- \delta_{s,s'+\hat{\nu}'}\delta_{\mu\mu'}
- \delta_{s,s'}\delta_{\mu\nu'}).$
When all $\tilde{f}_i$ are equal to $ f_i $,
one can prove an equality   
$ \langle \bar{S}_i \rangle  =  \langle S_i \rangle $,
where the expectation values are taken 
over the above original action 
with the coupling constants $\{f_i\}$. 
Otherwise, one may expand the difference as follows: 
\begin{equation}
\langle \bar{S}_i - S_i \rangle = 
\sum_j 
\langle \overline{S_i S_j}-\bar{S}_i\bar{S}_j \rangle
(f_j - \tilde{f}_j)    \label{ssbar}, 
\end{equation}
where only the first order terms are written down.
This allows an iteration scheme for determination 
of the unknown constants
$f_i$. 

Practically we have to restrict the number of 
interaction terms\cite{comment1}. We  adopted
12 types of quadratic interactions listed in Table \ref{table1} 
in most of these 
studies\cite{comment0}. 

First we applied the method to the Villain 
form of $U(1)$ lattice gauge 
theory\cite{villain,comment15}, since the Villain model 
is reformulated exactly as a theory of monopole 
loops having an action
\begin{equation}
\sum_{i} f_i S_i [k]= 2\pi^2\beta_v\sum_{s,s',\mu} k_{\mu}(s)
D(s-s')k_{\mu}(s') \label{vil},
\end{equation}
where $D(s)$ is the lattice Coulomb Green function\cite{bank}. 

Using the Villain action, we generated 100 gauge field configurations 
separated by 40 sweeps after a thermalization of 4000 sweeps at 
$\beta_v =0.60, 0.64$ and $0.68$ on $8^4$ lattice. 
We used the DeGrand-Toussaint scheme\cite{degrand} for locating 
monopole currents in the lattice gauge field configurations to obtain 
an ensemble of monopole configurations. 
The statistical errors were estimated 
with the jackknife method. 

The coupling constants $f_1 \sim f_5$ determined 
are plotted in Fig.1 in 
comprison with the theoretical values. 
$f_1$ agree well with those of the theoretical values, whereas 
there are small discrepancies with respect to $f_2 \sim f_5$. 
The discrepancies come from the truncation of the terms of the action 
taken. Really, if we include more quadratic terms, they disappear.

Next we applied the method to $SU(2)$ 
lattice gauge theory. Monte-Carlo simulations 
were done on $12^4, 14^4, 16^4, 
18^4, 20^4$ and $24^4$ lattices for 
$2.3 \le \beta \le 3.0$. The $U(1)$ link fields  are extracted 
after the abelian projection in the MA gauge. 
The DeGrand-Toussaint scheme
\cite{degrand} is used to get monopole configurations. 
Long-distance behaviors are expected to be important in the 
confinement phase of QCD. 
Hence we consider extended monopoles of the type-2
\cite{ivanenko,comment2,comment3}. 
The $n^3$ extended monopole of the type-2 has a total magnetic 
charge inside the $n^3$ cube and is defined on 
a sublattice with the spacing
$b=na$,  $a$ being the spacing of the original lattice\cite{comment4}.
The definition of the type-2 extended monopoles 
corresponds to making a block
spin transformation of the monopole currents with the scale factor $n$.
We call the sublattice as a renormalized lattice.
We derived the coupling constants for $1^3, 2^3, 3^3$ and $4^3$ 
extended monopoles.

Our results are summarized as follows:

1)The coupling constants $f_i$ are fixed for 
not so large value of
$b$ ,i.e., $b < 1.3\times 10^{-2}(\Lambda_L)^{-1}$. 
The iteration does not converge in the case
with  larger $b$. 
The coupling constants determined are almost 
independent of the lattice
volume as seen from Fig.2. 
We see $f_1$ is dominant and the coupling constants
decrease rapidly as the distance between 
the two monopole currents increases.

2)The energy of a monopole loop (with a unit charge
$|k_{\mu}(s)|=1$ \cite{comment45}) of length $L$ may be approximated 
by a self-energy part 
$f_1 L$, whereas its entropy grows like $L$ln7 
when $L$ is large\cite{bank}.
If $f_1 <$ ln7, the entropy dominates over the energy 
and the monopole condensation occurs.
We plot $f_1$ versus $\beta$ for various 
extended monopoles  on $24^4$ lattice
in comparison with the entropy value ln7 
for the infinite volume in Fig.3.
Each extended monopole has its own $\beta$ region where 
the condition  $f_1 <$ ln7 is satisfied. When  the extendedness 
is bigger, larger $\beta$ is included in such a region. 

3)It is interesting to see the relation between 
the monopole condensation and 
the deconfinement transition in a finite volume. In ref.\cite{kovac}, 
Polyakov loops are measured on symmetric lattices and the value 
$T_c =(25.8\pm1.1)\Lambda_L$ is given for the critical 'temperature'.
The critical coupling $\beta_c$ is fixied from 
the value for each lattice 
volume. For example, $\beta_c$ is $2.60$ 
on $14^4$ and $2.81$ on $24^4$ 
lattices, respectively.   

Our data indicate 
the following features irrespective of the original lattice volume.
If the renormalized lattice is larger than $7^4$, 
the $f_1$ value is above the ln7 line at $\beta_c$. 
It seems to cross the 
ln7 line just at $\beta_c$ when the 
renormalized lattice is about $7^4$. 
See Fig.2 for $2^3$ extended monopoles at $\beta=2.6$ 
which is just $\beta_c$ of $14^4$ lattice and in which 
$f_1$ takes the value about ln7.
Also ,in the Villain case shown in Fig.1, 
the energy-entropy arguments 
give the correct critical coupling on a lattice as small as $8^4$.
We may assert the above energy-entropy arguments can be used when 
the renormalized lattice is larger than $7^4$. 
Then we see  condensation of some extended 
monopoles occurs in the confinement phase.

On the other hand, when the renormalized 
lattice is smaller than $7^4$, 
such  simple arguments may not apply as seen from small 
discrepancy between 
the ln7 cross point of $f_1$ in the $4^3$ case and $\beta_c$ in Fig.3.
There is a possible entropy decreasing effect due to the periodic 
boundary condition on a small lattice. 
Such entropy calculations are very important to know 
if the confining phase and
the monopole-condensed phase are exactly 
the same as expected\cite{comment5}.

4)The behaviors of the coupling constants are 
different for different extended 
monopoles. But if we plot them versus $b$ , 
we get a unique curve as in Fig.4. The 
coupling constants seem to depend only on $b$, not on the extendedness 
nor $\beta$.  This suggests the existence 
of the continuum limit and  the monopole action 
in the limit may be similar to that given here.

5)From Fig.4, we see the self-energy per monopole loop length $f_1$ 
on the renormalized lattice decreases as 
the length scale $b$ increases.
A critical length $b_c \sim 5.2\times 10^{-3}(\Lambda_L)^{-1}$ exists 
at which the $f_1$ value crosses the ln7 line.
The $b$ dependence and the lattice-size independence of $f_1$ 
suggest the following picture  of the QCD 
vacuum on the standpoint of the monopole condensation.
When the physical volume of the system 
$l^4$ ($l=Na(\beta)$ on $N^4$ lattice)
is large enough, that is, $l\gg b_c$, 
one can take a large renormalized 
lattice with a spacing $b > b_c$. 
The entropy of monopole loops dominates 
the energy on such a renormalized lattice 
and the monopole condensation 
occurs. On the other hand, when the physical volume is small, 
the renormalized lattice must have a small lattice size 
or a small lattice spacing or both. A small 
lattice size would lead to 
a possible entropy decreasing  and a small 
spacing makes the self-energy large.
Thus the monopole condensation does not occur 
in a small physical volume.
The transition from the monopole-condensed phase to the normal phase 
takes place on a finite physical 
volume $l_c^4$. $l_c$ must be several 
times of $b_c$. If it corresponds to the deconfining phase 
transition, we get $l_c \sim 7b_c$.
The existence of $l_c$ means that there are 
always both monopole-condensed 
and uncondensed phases on a finite (in a lattice unit) lattice.

6) The monopole action may be fitted by
\begin{eqnarray*}
S & = & \sum m_0 b k_{\mu}(s)k_{\mu}(s) \\
&   &  + 
\frac{1}{2}(\frac{4\pi}{g(b)})^2 \sum 
k_{\mu}(s)\bar{D}(s-s')k_{\mu}(s'),
\end{eqnarray*}
where $g(b)$ is the SU(2) running coupling constant
\begin{eqnarray}
g(b)^{-2}= \frac{11}{24\pi^2}\ln(\frac{1}{b^2\Lambda^2}) 
+ \frac{17}{44\pi^2}
\ln\ln(\frac{1}{b^2\Lambda^2}).
\end{eqnarray}
The scale parameter determined is 
$\Lambda \sim 42\Lambda_L$\cite{comment6}.
$\bar{D}(s)$ is a 
modified lattice Coulomb propagator\cite{smit}.
For details, see ref.\cite{shiba3}. 
The solid line is the prediction given 
by the action with the parameters 
written in the figure.
This form of the action is predicted 
theoretically by Smit and Sijs\cite{smit}.
The existence of the bare monopole mass 
$m_0$ and the running coupling 
constant $g(b)$ is characterisitic of the 
action in comparison with that of 
compact QED.

In summary, QCD confinement may be interpreted as the dual Meissner
effect due to monopole condensation in the maximally abelian 
gauge. Our data is the first that shows the 
possible occurence of the monopole 
condensation in QCD. In the second part of this note\cite{shiba2}, 
it will be shown that the monopoles alone are 
responsible for the string 
tension also in $SU(2)$ QCD as is in compact QED\cite{stack}.
Details will be published elsewhere\cite{shiba3}.

We wish to acknowledge Yoshimi Matsubara for 
useful discussions especially on
entropy decreasing effects of monopole loops.
This work is financially supported by JSPS Grant-in Aid for 
Scientific  Research (c)(No.04640289).

\begin{table}
\caption{
The form of monopole action adopted. In addition to the following, 
all terms in which 
the relation of the two currents is equivalent are added. 
\label{table1}}
\begin{tabular}{rcr}
      i   &    interaction               &     \\
\tableline
      1   & $k_1(0,0,0,0)^2 $            &     \\
      2   & $k_1(0,0,0,0)k_1(1,0,0,0)$   &     \\
      3   & $k_1(0,0,0,0)k_1(0,1,0,0)$   &     \\
      4   & $k_1(0,0,0,0)k_1(1,1,0,0)$   &     \\
      5   & $k_1(0,0,0,0)k_1(0,1,1,0)$   &     \\
      6   & $k_1(0,0,0,0)k_1(2,0,0,0)$   &     \\
      7   & $k_1(0,0,0,0)k_1(1,1,1,0)$   &     \\
      8   & $k_1(0,0,0,0)k_1(0,1,1,1)$   &     \\
      9   & $k_1(0,0,0,0)k_1(1,1,1,1)$   &     \\
     10   & $k_1(0,0,0,0)k_1(2,1,0,0)$   &     \\
     11   & $k_1(0,0,0,0)k_2(2,1,0,0)$   &     \\
     12   & $k_1(0,0,0,0)k_1(2,1,1,0)$   &     \\
\end{tabular}
\end{table}

\begin{figure}[tb]
\caption{
Coupling constants $f_i$ versus $\beta$ in the Villain model. 
$f_1$ (circle), $f_2$ (square), $f_3$ (diamond), 
$f_4$ (triangle) and 
$f_5$ (cross) are plotted. Solid lines denote the theoretical values.
}
\end{figure}

\begin{figure}[tb]
\caption{Coupling constants $f_i$ versus lattice size . 
$f_1$ (circle), $f_2$ (square), $f_3$ (diamond), 
$f_4$ (triangle) and 
$f_5$ (cross) are plotted.
}
\end{figure}

\begin{figure}[tb]
\caption{Coupling constants $f_i$ versus $\beta$ for 
$2^3,3^3$ and $4^3$ 
extended monopoles on $24^4$ lattice. 
}
\end{figure}

\begin{figure}[tb]
\caption{Coupling constants $f_1$ versus $b$ }
\end{figure}


\begin{references}
\bibitem{comment1} 
All possible types of interactions are not independent, since 
$\partial'_{\mu}k_{\mu}(s)=0$. We can get rid of almost 
all interactions between 
different components of the currents from the quadratic action 
by use of the conservation rule.
\bibitem{comment0} To check the validity of the truncation, 
we  also included more quadratic
and even quartic interactions in some cases, but 
found little difference especially in $SU(2)$ QCD.
\bibitem{comment15}
We also applied the method to the Wilson form of  $U(1)$ lattice gauge 
theory. We obtained an action which is in approximate agreement with 
Eq.\ref{vil} with $\beta_v$ 
determined from the Wilson coupling constant
$\beta_w$ as shown in \cite{villain}.
\bibitem{comment2} Type-1 extended 
monopoles\cite{ivanenko} are not adopted 
here, since the separation of the Dirac string 
is problematic as shown in 
\cite{shiba3}.
\bibitem{comment3} We studied extended 
monopoles also in compact QED, but 
the iteration does not converge then.
\bibitem{comment4} Note that $b$ represents 
a scale at which behaviors of the 
monopole currents are examined. One should 
not think it to be the physical 
size of the monopoles, if any.
\bibitem{comment45} 
We have made a histogram analysis of the 
magnetic charge distribution 
of each extended monopole. The case 
with $|k_{\mu}(s)|= 2$ is much 
(less than 5\%) suppressed in comparison 
with the $|k_{\mu}(s)|=1$ case 
and charges with $|k_{\mu}(s)|\geq 3$ barely appear even 
for $4^3$ extended monopoles in the confinement 
phase. See \cite{shiba3}.
\bibitem{comment5} 
Such an entropy decreasing effect is actually seen when restricted 
(in its extreme limit, self-avoiding) 
random walks are considered on a small volume 
with periodic boundary conditions.For details, see \cite{shiba3}.
\bibitem{comment6}
We used also the data of $f_2$ for the fit. 
The form of the action reproduces 
also $f_3 \sim f_5$, but their experimental errors are too 
large to derive a definite 
conclusion.
\end{references}
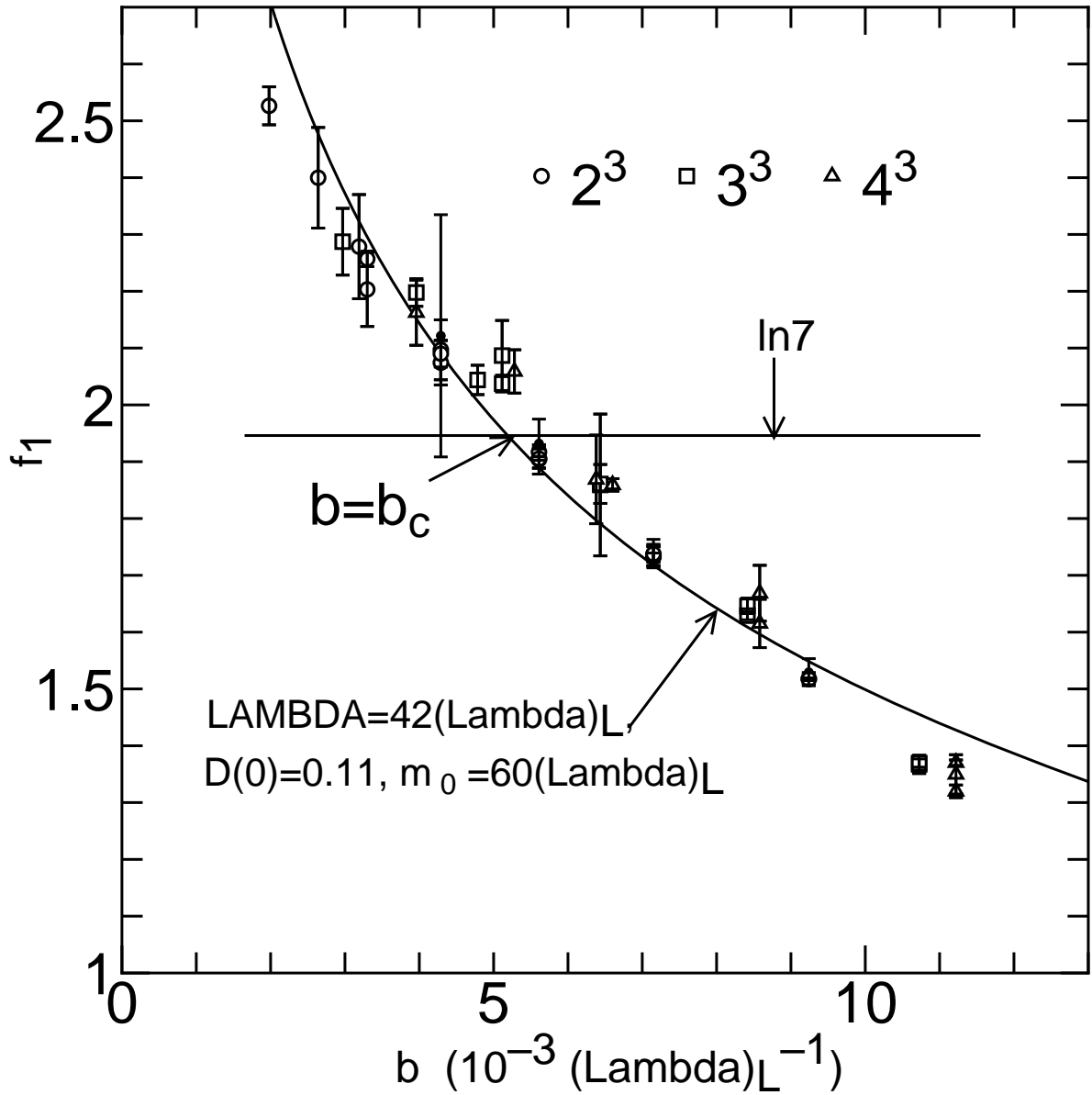
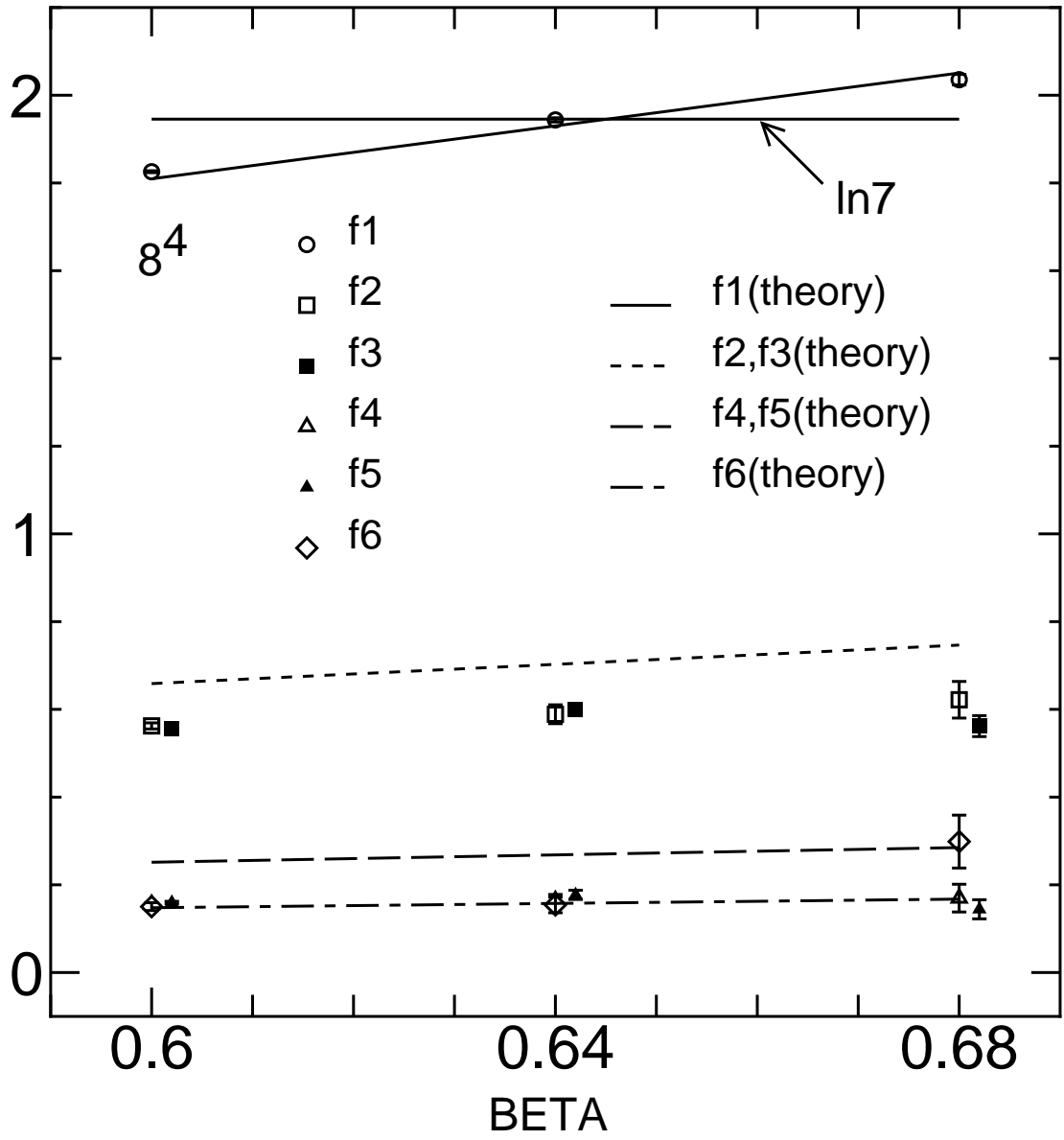
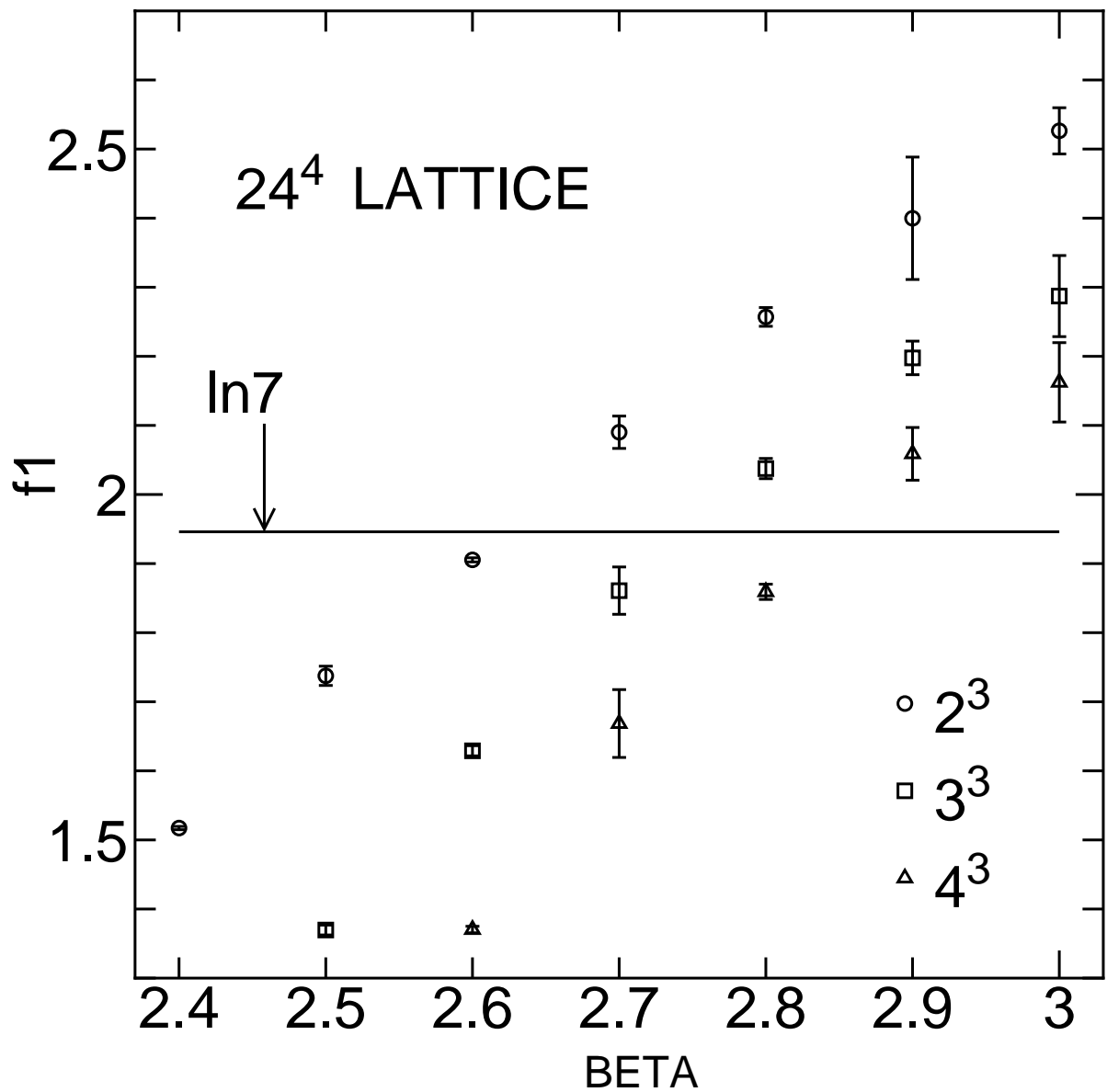
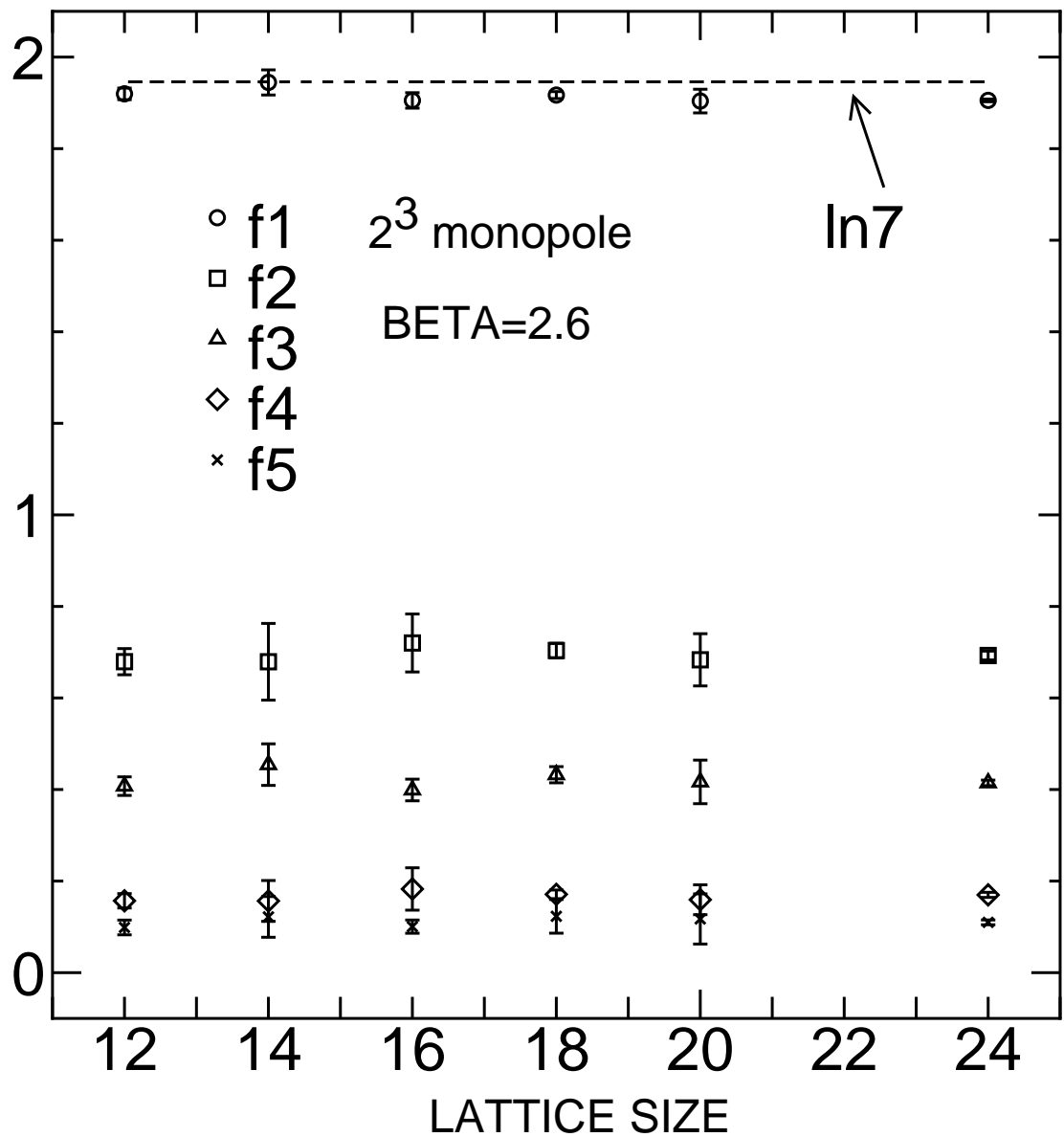
\end{document}